\documentclass[acmsmall,screen]{acmart}
\usepackage{url}

\usepackage[many]{tcolorbox}
\usepackage{nicefrac}
\usepackage{siunitx}
\usepackage{array,framed}
\usepackage{pdfpages}
\usepackage{booktabs}
\usepackage[utf8]{inputenc}  
\usepackage{
  color,
  float,
  epsfig,
  wrapfig,
  graphics,
  graphicx,
  subcaption
}
\usepackage{graphicx}
\usepackage{nicematrix}
\usepackage{wasysym}
\usepackage{pifont}
\usepackage{hyperref}
\usepackage{setspace}
\usepackage{amsfonts}
\usepackage{latexsym,fancyhdr}
\usepackage{enumerate}
\usepackage{fancybox}
\usepackage{fancyvrb}

\usepackage[linesnumbered,ruled,longend,noend]{algorithm2e}
\usepackage{algpseudocode}
\usepackage{graphicx}
\usepackage{xparse} 
\usepackage{xspace}
\usepackage{multirow}
\usepackage{csvsimple}
\usepackage{balance}
\usepackage{flushend}
\usepackage{pifont}
\usepackage{threeparttable}
\usepackage{longtable}
\usepackage{wrapfig}
\usepackage{shortcuts}
\usepackage{adjustbox}
\usepackage{booktabs} 
\usepackage{soul}
\usepackage{todonotes}
\usepackage{enumitem}
\usepackage{array}
\usepackage{color}
\usepackage{mdframed}
\usepackage{hyperref}
\usepackage{xcolor}
\usepackage{colortbl}
\usepackage{pdflscape}
\usepackage{tabularx}

\AtBeginDocument{%
  }


\author{Junzhe Yu}
\affiliation{%
  \institution{ShanghaiTech University}
  \city{Shanghai}
  \country{China}
}
\email{yujzh1@shanghaitech.edu.cn}

\author{Yi Liu}
\affiliation{%
  \institution{Quantstamp}
  \country{Singapore}
}
\email{yi009@e.ntu.edu.sg}

\author{Huijia Sun}
\affiliation{%
  \institution{ShanghaiTech University}
  \city{Shanghai}
  \country{China}
}
\email{sunhj2022@shanghaitech.edu.cn}

\author{Ling Shi}
\affiliation{%
  \institution{Nanyang Technological University}
  \country{Singapore}
}
\email{ling.shi@ntu.edu.sg}

\author{Yuqi Chen}
\authornote{Corresponding author.}
\affiliation{%
  \institution{ShanghaiTech University}
  \city{Shanghai}
  \country{China}
}
\email{chenyq@shanghaitech.edu.cn}

\begin{document}
\begin{abstract}
Large Language Models (LLMs) have significantly advanced text understanding and generation, becoming integral to applications across education, software development, healthcare, entertainment, and legal services. Despite considerable progress in improving model reliability, latency remains under-explored, particularly through recurrent generation—where models repeatedly produce similar or identical outputs, causing increased latency and potential Denial-of-Service (DoS) vulnerabilities.

We propose \toolgeneration{}, a black-box evolutionary algorithm that efficiently identifies recurrent generation scenarios in prominent LLMs like LLama-3 and GPT-4o. Additionally, we introduce \tooldetection{}, a lightweight real-time classifier trained on activation patterns, achieving 95.24\% accuracy and an F1 score of 0.87 in detecting recurrent loops. Our methods provide practical solutions to mitigate latency-related vulnerabilities, and we publicly share our tools and data to support further research.

\end{abstract}

\title{Breaking the Loop: Detecting and Mitigating Denial-of-Service Vulnerabilities in Large Language Models}
\maketitle

\section{Introduction}
Large Language Models (LLMs) have made significant advancements in understanding and generating human-like text, largely due to training on vast corpora of text. These models are increasingly being integrated into a wide range of software applications, spanning fields such as education~\cite{zhang2024simulating}, software development~\cite{guo2024stopefficientcodegeneration}, healthcare~\cite{restrepo2024analyzing}, entertainment~\cite{Perplexi34:online}, and legal services~\cite{izzidien2024llm}. For example, GRU~\cite{Assistan30:online} leverages LLMs to automatically generate code to resolve GitHub issues, achieving top rankings in the coding benchmark SWE-bench~\cite{jimenez2024swebenchlanguagemodelsresolve}. In these applications, interactions with LLMs—whether through offline or online systems—are essential for delivering contextually relevant responses. As a result, the quality and responsiveness of LLM-generated outputs are critical to the effectiveness of these applications.

While various aspects of LLM reliability have been explored within the software engineering community (e.g., quality of code generation~\cite{li2024cosec}, handling glitch tokens~\cite{10.1145/3660799}, and reliability of LLM embeddings~\cite{10.1145/3643767}), less attention has been given to an equally important factor—latency. Several factors such as the LLM type and configuration, hardware, and network conditions can influence response time in applications leveraging LLMs. However, in this work, we highlight an under-explored phenomenon known as \textit{recurrent generation} in LLMs, which leads to unexpectedly long latency across mainstream models, independent of these external factors.

\begin{figure}[t!]
\centering
\includegraphics[width=\linewidth]{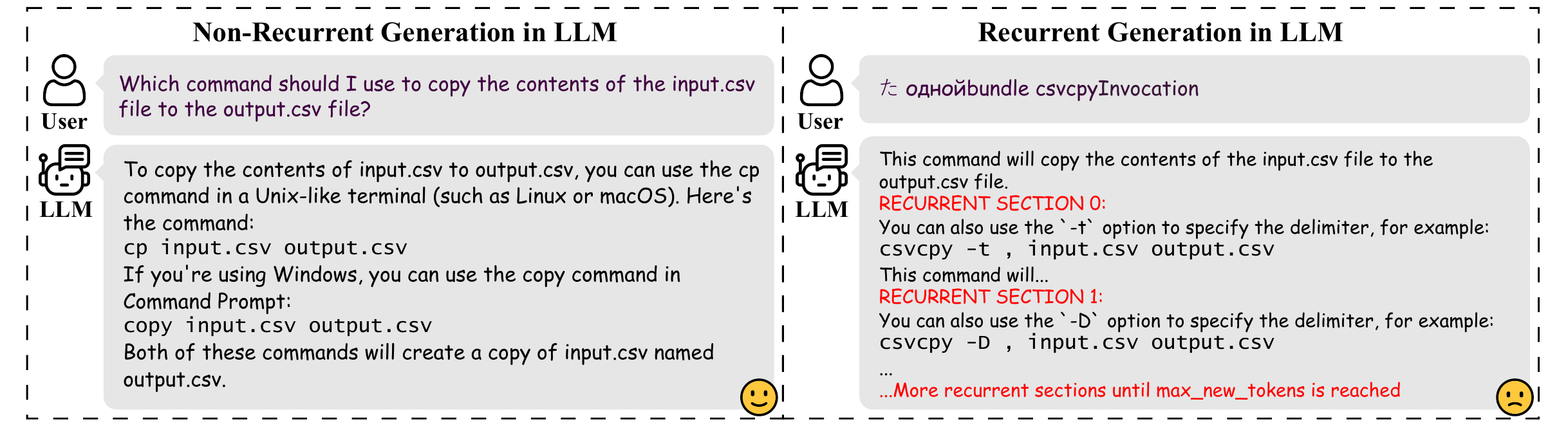}
\caption{Comparison between normal and recurrent generation in Llama2-7b-chat.}
\label{fig:demo}
\end{figure}
Recurrent generation occurs when LLMs repeatedly produce highly similar content, similar to how traditional infinite loops operate in programming~\cite{xie2017loopster,xie2016proteus}. As illustrated in Figure~\ref{fig:demo}, when a user inputs a prompt with random characters into Llama2-7b-chat~\cite{llama2}, the model continuously generates explanations for argument usage until it reaches the \textit{max\_new\_tokens} limit, which defines the maximum number of tokens the model can generate. This recurrent generation phenomenon affects users, developers, and LLM service providers. Recurrent generation significantly extends latency, deteriorates user experience, and introduces the risk of Denial-of-Service (DoS) attacks. From the users' perspective, recurrent generation leads to repetitive content, causing long wait times and diminishing the overall user experience in applications integrated with LLMs. For developers, who typically pay for LLM services based on the number of tokens consumed (e.g., in GPT-4o~\cite{PricingO96:online}), detecting and mitigating recurrent generation can enhance user satisfaction and reduce costs by preventing the generation of redundant content. For LLM service providers like OpenAI~\cite{OpenAI3:online}, addressing recurrent generation can significantly improve the throughput of their services by reducing the length of token sequences, thereby lowering hardware usage (e.g., GPU utilization) and saving energy.

However, existing work on studying recurrent generation in LLMs is limited. The most relevant concept is the ``sponge example''~\cite{sponge-example-1} from the pre-LLM era, where crafted text inputs caused latency degradation in neural networks, such as recurrent networks. The concept of the sponge example was first defined by Ilia et al.~\cite{sponge-example-1}, who proposed a gradient-based white-box methodology to generate sponge examples that degraded the performance of systems like Microsoft Translator. Several follow-up studies optimized the efficiency of finding sponge examples~\cite{cinà2023energylatencyattacksspongepoisoning,shapira2023phantom,deng2024deconstructing} using white-box approaches. While effective for small models, generating samples that exhibit recurrent generation is particularly challenging in white-box settings for LLMs. Existing sponge example methods~\cite{sponge-example-1,cinà2023energylatencyattacksspongepoisoning} rely on gradient-based approaches that require instrumenting the LLM under test, which introduces significant runtime overhead. Additionally, gradient-based optimization itself is time-consuming. Furthermore, some LLMs, such as GPT-4o~\cite{openai2023gpt4}, do not provide access to internal model states, making white-box approaches infeasible. Consequently, testing LLMs for this aspect presents a significant challenge. Additionally, detecting recurrent generation in LLMs introduces its own set of difficulties, as the underlying causes of recurrent generation remain unclear, complicating the design of effective and efficient detection methods.

To tackle these challenges, we first design a black-box generation method for creating recurrent generation samples across various LLMs. Our goal is to demonstrate that this phenomenon is prevalent across different models and to efficiently generate a benchmark for the detection phase. Next, by investigating the root causes of recurrent generation, we develop a detection method to address the phenomenon effectively.

Specifically, in this work, we aim to explore the following research questions:

\noindent$\bullet$ \textbf{RQ1 (Recurrent Generation): How efficiently can recurrent generation be triggered in LLMs?}

Efficiently generating test inputs that trigger recurrent generation can help developers and LLM service providers evaluate their applications. It also provides datasets for analyzing, detecting, and mitigating recurrent generation in LLMs. In this work, we propose \toolgeneration{}, an evolutionary algorithm designed to generate such test inputs effectively. Ultimately, we identified a total of 2,388 test inputs that trigger recurrent generation in eight top LLMs, including LLama-3~\cite{llama3} and GPT-4o~\cite{gpt4}.

\noindent$\bullet$ \textbf{RQ2 (Characteristics): What are the characteristics of recurrent generation across different LLMs?}

This research question aims to provide insights into how LLMs behave internally when encountering recurrent generation. We analyze the dataset collected from RQ1 and discover that recurrent generation tends to exhibit similar activation patterns across different models.

\noindent$\bullet$ \textbf{RQ3 (Detection): How effectively can we detect recurrent generation across LLMs?}

Based on the insights gained from RQ2, we propose \tooldetection{}, a lightweight multi-layer perceptron (MLP) trained on activation patterns to detect recurrent generation in LLMs in real-time.

\textbf{Contributions.} We summarize our key contributions as follows:

\begin{itemize}
    \item \textbf{Recurrent Generation Test Inputs.} We have developed an efficient black-box evolutionary algorithm, \toolgeneration{}, to optimize and generate test inputs that trigger recurrent generation in eight top LLMs like GPT-4o and GPT-4o mini. In total, we have identified 2,388 test inputs that cause recurrent generation in these models, including LLama-3~\cite{llama3} and GPT-4o~\cite{gpt4}.
    
    \item \textbf{New Findings.} Based on the generated test inputs, we observed that recurrent generation in LLMs is associated with specific patterns in the models' activations and hidden states when they become trapped in a recurrent loop.
    
    \item \textbf{Recurrent Generation Detection.} Building on the insights obtained, and to the best of our knowledge, we propose the first practical detection technique, \tooldetection{}, an MLP classifier trained on activation patterns from the recurrent generation dataset. We achieved high accuracy (95.2\%), a high F1 score (0.87), and a low false positive rate (2.6\%) in six top open-source LLMs like LLama-3 and Gemma-2.
    
    \item \textbf{Open Source Artifact.} We release the code and results of our experiments on our website~\cite{our-website}, providing resources to support and encourage further research on latency-related vulnerabilities and DoS mitigation strategies for LLM-based systems.
\end{itemize}
\section{Background}

\subsection{Token and Tokenization in Large Language Models}

In the field of natural language processing (NLP), \textit{tokenization} is the process of breaking down raw text into smaller units called \textit{tokens}~\cite{vijayarani2016text}. A token, denoted as $t_i$, can represent a word, subword, or character, depending on the granularity of the tokenization process. For example, given the sentence \textit{``The cat sat on the mat''}, a simple tokenization would produce the following sequence of word tokens:
\[
[t_1, t_2, t_3, t_4, t_5, t_6] = [\text{``The''}, \text{``cat''}, \text{``sat''}, \text{``on''}, \text{``the''}, \text{``mat''}].
\]
In more complex cases, such as subword tokenization~\cite{tay2021charformer}, a word like \textit{``unbelievable''} could be broken down into subwords as:
\[
[t_1, t_2, t_3] = [\text{``un''}, \text{``believ''}, \text{``able''}],
\]
which allows for better generalization across similar tokens (e.g., \textit{``belief''}, \textit{``believer''}). All tokens produced during tokenization form a \textit{vocabulary}, $V$, which is a finite set of unique tokens that the model can process. This vocabulary is essential for transforming text into numerical representations that can be understood and manipulated by the model. 

In the context of LLMs, such as GPT-4o~\cite{gpt4}, tokenization is a crucial preprocessing step. Let the input text $X$ be a sequence of tokens $[t_1, t_2, \dots, t_n]$. The model processes these tokens to predict the next token in the sequence, aiming to generate the most probable token $t_{n+1}$ given the previous tokens. For example, given the tokenized input $[t_1, t_2, t_3, t_4] = [\text{``The''}, \text{``cat''}, \text{``sat''}, \text{``on''}]$, the model would predict the next token $t_5 = \text{``the''}$, followed by $t_6 = \text{``mat''}$.

Tokenization not only enables efficient processing of text in LLMs but also aids in managing memory and computational efficiency~\cite{tay2021charformer}, as the model operates on sequences of tokens rather than full sentences or paragraphs. This technique is fundamental for enabling LLMs to perform a wide range of tasks, including text generation~\cite{Assistan30:online}, translation~\cite{yao2023fuzzllm}, summarization~\cite{vijayarani2016text}, and more.

\subsection{Recurrent Generation in Large Language Models}

Recurrent generation refers to a phenomenon in which an LLM generates responses that are highly similar but not necessarily identical, eventually hitting the maximum output token length limit during a single generative process. Formally, let $\mathcal{M}$ represent an LLM, and let $T = [t_1, t_2, \dots, t_n]$ be a sequence of tokens generated by $\mathcal{M}$. Recurrent generation occurs when a subsequence $S = [t_{i}, t_{i+1}, \dots, t_{i+k}]$ is repeated with slight variations, producing a new subsequence $S' = [t_{j}, t_{j+1}, \dots, t_{j+k}]$ where $j > i+k$. The content of $S'$ is highly similar to $S$, with minor differences such as changes in variables, filenames, or other contextual elements.

LLMs use a parameter, \texttt{max\_new\_tokens}, to control the maximum number of tokens that can be generated in one response. In this work, we consider recurrent generation valid only when the LLM reaches the \texttt{max\_new\_tokens} limit. For example, as illustrated in Figure~\ref{fig:demo}, Llama2-7b-chat repeatedly generates redundant argument help usage, leading to redundancy without meaningful content variation. This behavior results in inefficient token usage, increased latency, and a diminished user experience. Addressing recurrent generation is crucial for improving the efficiency of LLM-based systems, as it reduces computational costs, optimizes token consumption, and enhances user satisfaction.

\section{\toolgeneration{}: Recurrent Generation in LLMs}
\label{sec:generation-methodology}
\begin{figure}[t!]
\centering
\includegraphics[width=\linewidth]{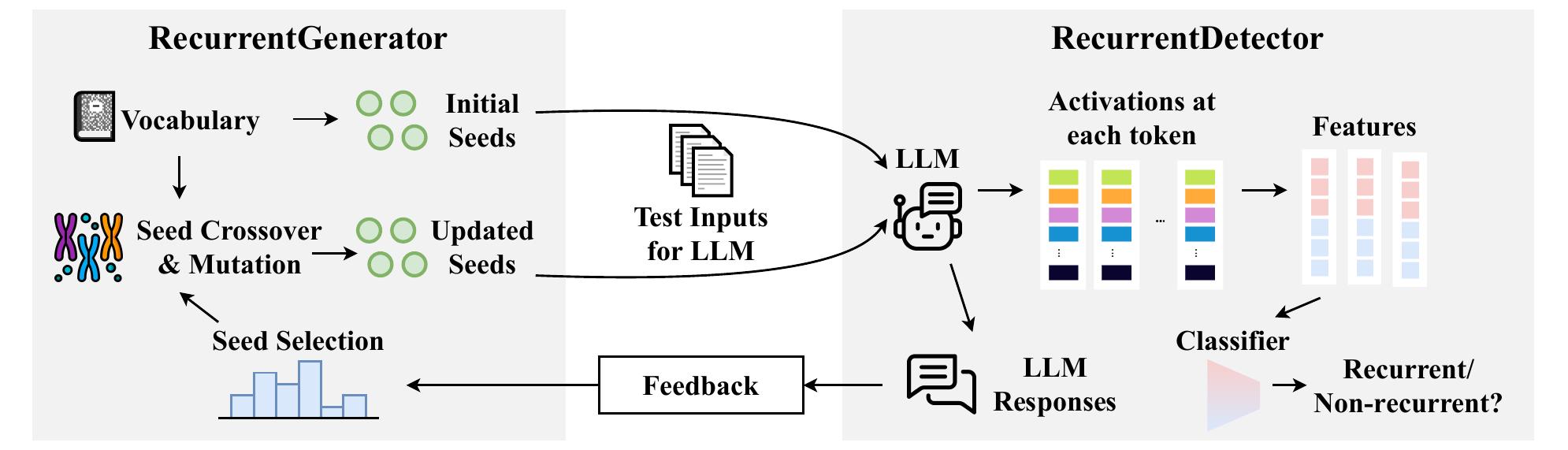}
\caption{Overview of \toolgeneration{} (\S{}~\ref{sec:generation-methodology}) and \tooldetection{} (\S{}~\ref{sec:detection-methodology}).}
\label{fig:overview}
\end{figure}

As illustrated in Figure~\ref{fig:overview}, we present the complete framework, which consists of two main components. The first component, \toolgeneration{}, is responsible for generating test inputs for LLMs that trigger recurrent generation. The second component, \tooldetection{}, builds upon the analysis of these test inputs to implement a lightweight, real-time detector that identifies recurrent generation during LLM output generation. Details of \tooldetection{} will be discussed in \S{}~\ref{sec:detection-methodology}. In this section, we focus specifically on the methodology used to generate test inputs that effectively trigger recurrent generation. 

\subsection{Preliminary Study}
\label{sec:generation-preliminary-study}
We begin by investigating how to generate test inputs that trigger recurrent generation in LLMs through a preliminary study.

In this study, we use a random sampling strategy, where $L$ distinct tokens are randomly selected from the LLM's token vocabulary $V$ and fed into the LLM under test to observe its output. We chose Llama2-7B-chat~\cite{llama2} as the model for evaluation and varied $n$ (the number of tokens) between 4 and 15 to assess the impact of token length on recurrent generation. Sampling is conducted up to a maximum of 10,000 attempts, with the temperature parameter set to 0 to ensure reproducibility. Additionally, we set \texttt{max\_new\_tokens} to 2,000. This limit was chosen based on an analysis of popular real-world LLM interaction datasets, SharGPT~\cite{sharegpt}, where over 99.7\% of responses are shorter than 2,000 tokens. This ensures that all detected recurrent generations are true positives, as the randomly sampled tokens are expected to produce nonsensical inputs that should not typically generate long responses. All other parameters for Llama2-7B-chat remain at their default values, and the experiment is repeated 10 times to mitigate the effects of randomness.

\begin{figure}[t!]
\centering
\includegraphics[width=0.7\linewidth]{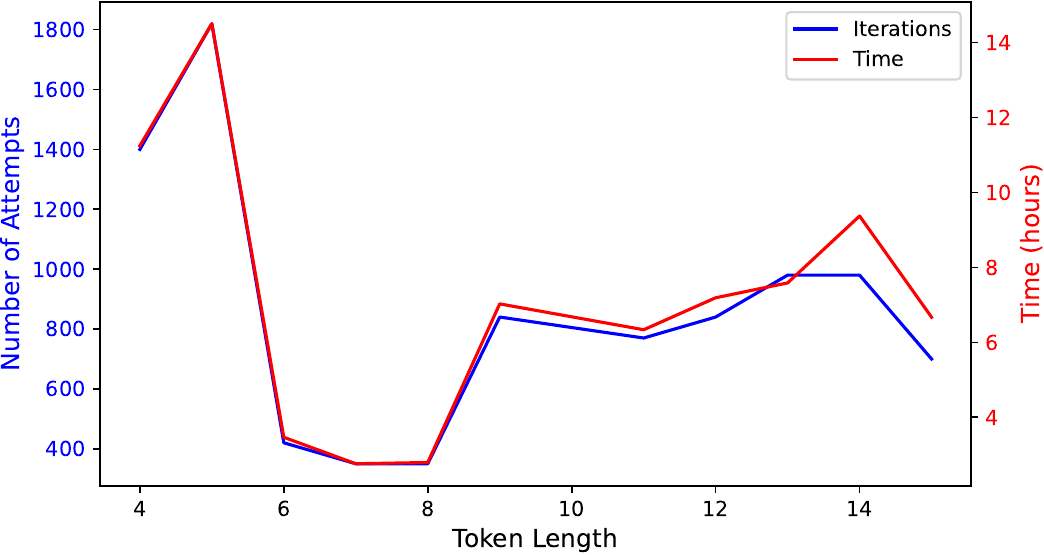}
\caption{Line chart illustrating the average number of attempts required and the total time cost to identify the first recurrent generation input across different token lengths in Llama-7b-chat. The chart reveals that a token length of 8 is optimal, minimizing both the number of attempts and the total time cost.}
\label{fig:preliminary-study-token-length}
\end{figure}

Figure~\ref{fig:preliminary-study-token-length} presents the average number of random attempts required to trigger recurrent generation for each token length in Llama2-7B-chat. The results indicate that recurrent generation is relatively easy to induce, with the maximum number of attempts being 1,803.2. Notably, for a token length of 8, only 389.5 sampling attempts are needed on average to find a valid test input that triggers recurrent generation.

To determine whether all responses that reach the \texttt{max\_new\_tokens} limit exhibit recurrent generation, we manually reviewed the outputs from Llama2-7B-chat. We have found that all such responses contained repeatedly generated content, differing mainly in the length of the repeated segments. For instance, a token sequence like ``auf extendedvariant fils Printpec laravel'' causes the model to generate a step-by-step explanation for creating a database table named ``laravel,'' with only the table name varying in each repetition. Similarly, the token ``radekieLinearLayout+ estadounidenseebru Pack convex'' results in the model repeating the sentence ``\texttt{LinearLayout} with a \texttt{ConvexShape} in Android'' multiple times.

We also evaluate whether benign inputs could trigger responses that hit the \texttt{max\_new\_tokens} limit. A manually provided prompt such as ``Tell me a long story'' generated a response of approximately 400 tokens. In contrast, a nonsensical prompt like ``ssl GUI servers tema otros'' caused the model to reach the 2,000-token limit during our evaluation.

Although random sampling can identify test inputs that trigger recurrent generation in LLMs, it still requires hundreds of attempts. This observation motivates us to explore more efficient methods for generating test inputs that can reliably trigger recurrent generation in LLMs.

\subsection{Our Approach}

We formulate the task of finding recurrent generation in LLMs as a search problem, analogous to traditional test case generation~\cite{fraser2011evolutionary,fraser2011evosuite,fraser2014large}. Given a token vocabulary $V$ from the LLM under test and a specified token length $L$, our objective is to identify as many test inputs of length $L$ as possible that lead to recurrent generation within a fixed number of iterations.

To efficiently search for such test inputs, we draw inspiration from evolutionary algorithms commonly used in test case generation~\cite{fraser2011evolutionary,fraser2011evosuite,fraser2014large}. We propose \toolgeneration{}, an evolutionary-based algorithm designed to generate test inputs that trigger recurrent generation in LLMs effectively. We select an evolutionary approach due to its ability to balance both diversity and effectiveness in the search process.

Specifically, evolutionary algorithms leverage a fitness function to evaluate the success of each individual test case, ensuring that offspring are increasingly effective. In our context, this means generating inputs that lead to longer and similar LLM responses exhibiting recurrent generation. To maintain diversity, these algorithms apply crossover and mutation techniques to combine and modify successful parent test cases, producing new offspring with enhanced qualities. This increases the likelihood of triggering recurrent generation in LLMs.

\subsection{Methodology Overview}

As shown in Algorithm~\ref{algo:ga_black_box}, \toolgeneration{} implements an evolutionary algorithm aimed at optimizing test inputs to trigger the longest possible text generation in an LLM. The algorithm begins by initializing a population of randomly sampled tokens of length \(L\) and iteratively improves this population over several iterations using selection, crossover, and mutation operations, which are detailed in the following sections.

More specifically, the \textbf{input} to the algorithm includes the large language model under test \(M\), its vocabulary \(V\), the input token length \(L\), the population size \(N\), the number of iterations \(T\), the selection rate \(r\), the number of fitness evaluations per prompt \(E\), and the mutation ratio \(m\). The \textbf{output} is a set of optimized text inputs \(P\) designed to trigger the longest possible recurrent generation from the LLM under test.

\begin{algorithm}[t!]
\caption{Algorithm of \toolgeneration{}}
\label{algo:ga_black_box}\label{something:ga_black_box}
\SetKwInOut{Input}{Input}
\SetKwInOut{Output}{Output}

\Input{Large Language Model under test $M$, its vocab $V$, input token length $L$, pool size $N$, number of iterations $T$, selection rate $r$, fitness evaluation times $E$, sequence mutation ratio $m$}
\Output{A set of test inputs triggering the recurrent generation, $P$}

\SetKwFunction{FMain}{Main}
\SetKwFunction{FRandomInit}{RandomInit}
\SetKwFunction{FSelect}{Select}
\SetKwFunction{FCrossOver}{CrossOver}
\SetKwFunction{FMutation}{RandomMutation}
\SetKwFunction{FSampleWithReplacement}{sampleWithReplacement}
\SetKwFunction{FSampleWithoutReplacement}{sampleWithoutReplacement}
\SetKwFunction{FAverage}{average}
\SetKwFunction{FRandomSample}{randomSample}
\SetKwFunction{FFitnessScore}{fitnessScore}
\SetKwFunction{FSelectBest}{selectBest}

\SetKwProg{Fn}{Function}{:}{}
\Fn{\FMain{$M, V, L, N, T, r, E, m$}}{
    $P \gets \FRandomInit{N, L, V}$\;
    \For{$i \gets 1$ \textbf{to} $T$}{
        $P \gets \FSelect{P, M, E, r}$\;
        \While{$|P| < N$}{
            $O \gets \FCrossOver{P, L}$\;
            $P \gets P \cup \FMutation{O, L, V, m}$\;
        }
    }
    \Return $P$\;
}

\Fn{\FRandomInit{$N, L, V$}}{
    $P \gets \{\}$\;
    \For{$i \gets 1$ \textbf{to} $N$}{
        \tcp{Take a random sample of $L$ tokens from vocab $V$}
        $P \gets P \cup \FRandomSample(L, V)$\;
    }
    \Return $P$\;
}

\Fn{\FSelect{$P, M, E, r$}}{
    $F \gets \{\}$\;
    \For{\textbf{each} $p \in P$}{
        $\text{response\_batch} \gets M(p, E)$\;
        $F \gets F \cup (p, \FAverage(\FFitnessScore(\text{response\_batch})))$\; 
    }
    \tcp{Select best candidates with ratio $r$}
    \Return $\FSelectBest(F, \lfloor |P| \times r \rfloor)$\; 
}

\Fn{\FCrossOver{$P, L$}}{
    \tcp{Taking two samples with replacement of the population $p$}
    $p, q \gets \FSampleWithReplacement(P, 2)$\; 
    $p_{\text{half}} \gets p[1 \ldots \lfloor L/2 \rfloor]$\;
    $q_{\text{half}} \gets q[\lfloor L/2 \rfloor + 1 \ldots L]$\;
    \Return $p_{\text{half}} + q_{\text{half}}$\;
}

\Fn{\FMutation{$O, L, V, m$}}{
    $I \gets \FSampleWithoutReplacement(\{1, \ldots L\}, \lceil |L| \times m \rceil)$\; \tcp{select a subset of size $\lceil |L| \times m \rceil$ of indexes to mutate}
    \For{\textbf{each} $i \in I$}{
        $O[i] \gets \FRandomSample(1, V)$\;
    }
    \Return $O$\;
}

\end{algorithm}

\subsection{Initialization of \toolgeneration{}}

At the beginning, \toolgeneration{} generates \(N\) random test inputs, each of length \(L\), by sampling tokens from the vocabulary \(V\). This forms the initial population of test inputs, which will be evolved in subsequent generations.

\subsection{Fitness Function and Population Selection}

In this section, we describe how effective test inputs are evaluated and retained for subsequent iterations. The \textbf{selection function} (\texttt{Select}) in Algorithm~\ref{algo:ga_black_box} evaluates the fitness of each test input based on the responses generated by the LLM \(M\). Since the LLM uses weighted sampling~\cite{APIRefer16:online} to choose the next token, we collect \(E\) responses for each input to reduce randomness, specifically measuring the length of the LLM's responses. The top \(\lfloor |P| \times r \rfloor\) test inputs, determined by their fitness scores, are selected through the \texttt{selectBest} function. This ensures that only the most effective inputs are used for breeding in the next iteration.

\noindent\textbf{Design of the Fitness Function.} As discussed in the preliminary study (\S~\ref{sec:generation-preliminary-study}), manual inspection of output sequences reveals that longer sequences often exhibit repetitive patterns, where the LLM generates redundant content. Based on this observation, we have designed a heuristic fitness function that incorporates a self-similarity score, calculated using the probability distributions of the output tokens:

$$
S(P) = \frac{1}{|\text{Pairs}(l_{\text{out}})|} \sum_{(i, j) \in \text{Pairs}(l_{\text{out}})} P_i \cdot P_j
$$

Where:

\begin{itemize}
    \item \(P\) is the sequence of probability distributions for the next tokens in the output, of length \(l_{\text{out}}\).
    \item \(\text{Pairs}(l_{\text{out}})\) represents all unique pairs of indices \((i, j)\) where \(0 \leq i < j \leq l_{\text{out}}\).
    \item \(P_i \cdot P_j\) denotes the inner product (dot product) of the probability distributions at positions \(i\) and \(j\).
    \item \(S(P)\) is the self-similarity score of the output sequence.
\end{itemize}

This score quantifies the average similarity between pairs of probability distributions within the output sequence, providing insight into the presence of recurrent patterns.

\begin{figure}[t!]
\centering
\includegraphics[width=0.8\linewidth]{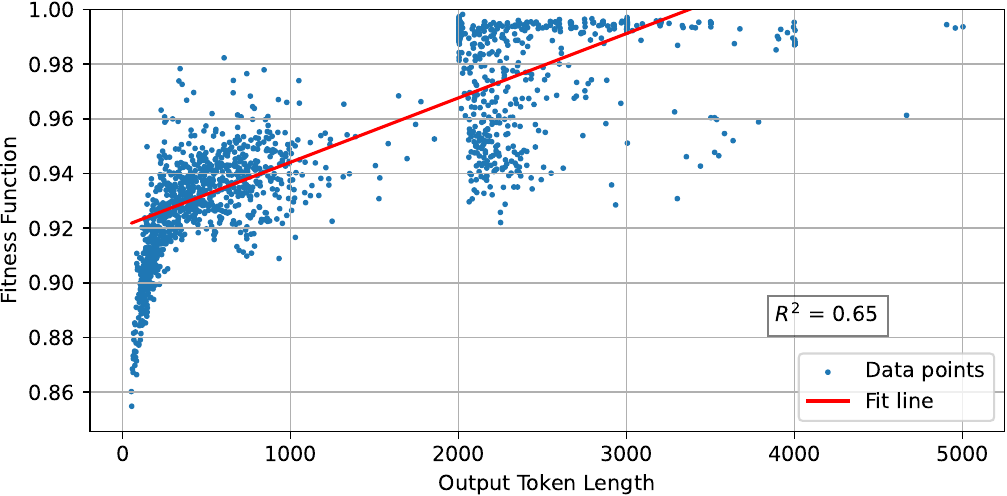}
\caption{Scatter plot and linear regression showing the correlation between the self-similarity fitness function and the response token length in Llama2-7b-chat.}
\label{fig:self_similarity_corr}
\end{figure}

Figure~\ref{fig:self_similarity_corr} illustrates the correlation between the self-similarity heuristic and the generated token length in Llama2-7b-chat. The scatter plot shows a clear positive correlation between the length of the generated sequences and their self-similarity scores, with $R^2 = 0.65$. Additionally, there is a notable gap in token length between normal outputs and those with excessive length. These observations support the rationale behind using self-similarity as a key component of the fitness function.

\subsection{Crossover and Mutation Function} 

The \textbf{crossover function} (\texttt{CrossOver}) generates new test inputs by combining segments of two selected test inputs. Specifically, it takes the first half of one test input and merges it with the second half of another to form a new test input. This process introduces new combinations of tokens, which may result in more effective inputs compared to the original ones.

The \textbf{mutation function} (\texttt{RandomMutation}) introduces variation into the new test inputs by randomly altering parts of the input. It selects specific positions within a test input and replaces them with new tokens sampled from the vocabulary \(V\), modifying up to $\lceil L \times m \rceil$ tokens. This step ensures the population retains diversity, preventing premature convergence on suboptimal solutions.

\vspace{-1em}
\section{Evaluation of \toolgeneration{}}
\label{sec:evaluation-generation}
In this section, we aim to evaluate the effectiveness of \toolgeneration{} and answer the following question: \textbf{RQ1 (Recurrent Generation)}: How efficiently can recurrent generation be triggered in LLMs?

\vspace{-1em}
\subsection{Experiment Setup}

\noindent\textbf{Baselines.} To the best of our knowledge, no existing baselines address recurrent generation in LLMs. To evaluate the effectiveness of \toolgeneration{}, we implement a random strategy that samples random token lengths $L$ over a fixed number of sampling attempts.

\noindent\textbf{LLMs under test.} We evaluate eight LLMs, including six popular open-source models and two commercial models. These models, encompassing various versions of LLama and Gemma, were chosen for their widespread use and strong performance in natural language processing tasks. The specific models tested are:

\begin{itemize}
    \item \textbf{LLama-3 (8B)}~\cite{llama3}: The latest iteration in the LLama series, offering significant improvements in both size and performance. The 8-billion-parameter model (8B) is tested to assess its performance in cutting-edge technologies.
    
    \item \textbf{LLama-2 (7B and 13B versions)}~\cite{llama2}: The second generation of LLama models, known for enhancements in both efficiency and accuracy.
    
    \item \textbf{LLama (7B and 13B versions)}~\cite{llama}: We test Vicuna-v1.5-7B~\cite{vicuna} and Vicuna-v1.5-13B~\cite{vicuna13}, which are fine-tuned versions of the original LLama models.
    
    \item \textbf{Gemma-2 (2B)}~\cite{gemmateam2024gemma2improvingopen}: We select the latest LLM developed by Google to evaluate the generalizability of \toolgeneration{} across different architectures.
    
    \item \textbf{GPT-4o and GPT-4o mini}~\cite{gpt4}: The most advanced commercial LLMs are included to assess the effectiveness of \toolgeneration{} in real-world applications.
\end{itemize}

\noindent\textbf{Experimental Settings.} All experiments were conducted on a system with four NVIDIA Titan RTX GPUs, each with 24\,GB of memory, running Ubuntu 22.04. To provide a fair comparison, the random baseline was configured with a total of 5,000 attempts, calculated as 50 (population size) $\times$ 20 (iterations) $\times$ 5 (evaluation repetitions), matching the maximum number of attempts used by \toolgeneration{}. The LLMs were configured according to their respective instructions with the temperature set to 0. We set the token length $L = 8$ in \toolgeneration{}, as Figure~\ref{fig:preliminary-study-token-length} indicates that this length requires the fewest attempts to generate a valid test input that triggers recurrent generation. We set the population size to 50 and ran the algorithm for 20 iterations, recording all test inputs that resulted in LLM responses reaching the \texttt{max\_token} limit of 2,000 tokens, which we consider as evidence of recurrent generation. For each input, we performed inference 5 times ($E = 5$) to estimate the expected fitness score. At each iteration, we selected the top 20\% of individuals ($r = 0.2$) based on their fitness scores. The offspring had 10\% of their tokens mutated ($m = 0.1$). To mitigate the effects of randomness, we repeated all experiments ten times.

\subsection{Evaluation Results}
\begin{table}[t!]
    \centering    
    \resizebox{\textwidth}{!}{
        \begin{tabular}{l|cccccccc|l} 
\toprule
\multirow{2}{*}{\textbf{Methods}} & \multicolumn{8}{c|}{\textbf{Models}}                                                                       & \multirow{2}{*}{\textbf{Average}}  \\
                                  & Gemma2-2B & Vicunna-v1.5-7B & Vicunna-v1.5-13B & Llama2-7B & Llama2-13B & Llama3-8B & GPT-4o-mini & GPT-4o &                                    \\ 
\hline
\textbf{RecurrentGenerator}       & 3.1       & 10.1            & 20.3             & 170.4     & 120.9      & 163.1     & 551.7       & 1136.8 & \multicolumn{1}{c}{272.1}          \\ 
Random                            & 4.7       & 20.4            & 71.6             & 684.3     & 1716.6     & 935.4     & Failed      & Failed & \multicolumn{1}{c}{1679.1}         \\
\bottomrule
\end{tabular}
    }
    \caption{Average number of attempts needed by \toolgeneration{} and a random generator to find recurrent samples. The random generator failed to find samples within 5000 attempts for both GPT-4o and GPT-4o-mini.}
    \label{tab:generation-average-attempts}
\end{table}

\noindent\textbf{Generation Efficiency across LLMs.} Table~\ref{tab:generation-average-attempts} presents the average number of attempts required to generate one test input that triggers recurrent generation in LLMs. We find that \toolgeneration{} can efficiently identify test inputs causing recurrent generation in an average of 272.1 attempts, which is significantly more efficient than the random strategy (a speed-up of 517\%), requiring an average of 1,679.1 attempts to generate a valid test input. Surprisingly, even the most advanced LLMs, such as GPT-4o and GPT-4o-mini, are susceptible to recurrent generation, requiring no more than 1,136.8 attempts. Gemma2-2B appears to be the most fragile LLM, requiring only 3.1 attempts to find a test input that triggers recurrent generation. Further investigation shows that as more effective test inputs evolve within the population, \toolgeneration{} continues to refine these inputs, generating increasingly valid test cases. This demonstrates that \toolgeneration{}, leveraging an evolutionary algorithm, is both effective and efficient at generating test inputs.

\begin{tcolorbox}[colback=gray!25!white, size=title,breakable,boxsep=1mm,colframe=white,before={\vskip1mm}, after={\vskip0mm}]
\textbf{Finding 1:} Even the most advanced LLMs, such as GPT-4o and LLaMa3-8B, are prone to recurrent generation and can be triggered with relatively few attempts.
\end{tcolorbox}

\noindent\textbf{Case Study.} We manually sampled and analyzed 800 test inputs and their corresponding responses from eight LLMs, with 100 samples from each LLM. We found that all LLM responses exhibited recurrent generation, repeating highly similar content until they reached the 4,000-token limit. For example, one test input, ``stormswwismsSQL Creation np 240 names,'' caused GPT-4o-mini to repeatedly generate SQL insertion commands. Similarly, the input ``AskicipLinearLayout+USTebru andfilters'' prompted Llama2-7B to repeatedly output the character ``ASpectRatioLayout.'' These findings indicate that recurrent generation is not only common in LLMs but also leads to redundant, nonsensical content that degrades model performance.

\begin{tcolorbox}[colback=gray!25!white, size=title,breakable,boxsep=1mm,colframe=white,before={\vskip1mm}, after={\vskip0mm}]
\noindent\textbf{Answer to RQ1:} \toolgeneration{} efficiently generates test inputs that trigger recurrent generation across various LLMs using an evolutionary algorithm.
\end{tcolorbox}

\begin{figure}[ht!]
\centering
\includegraphics[width=\linewidth]{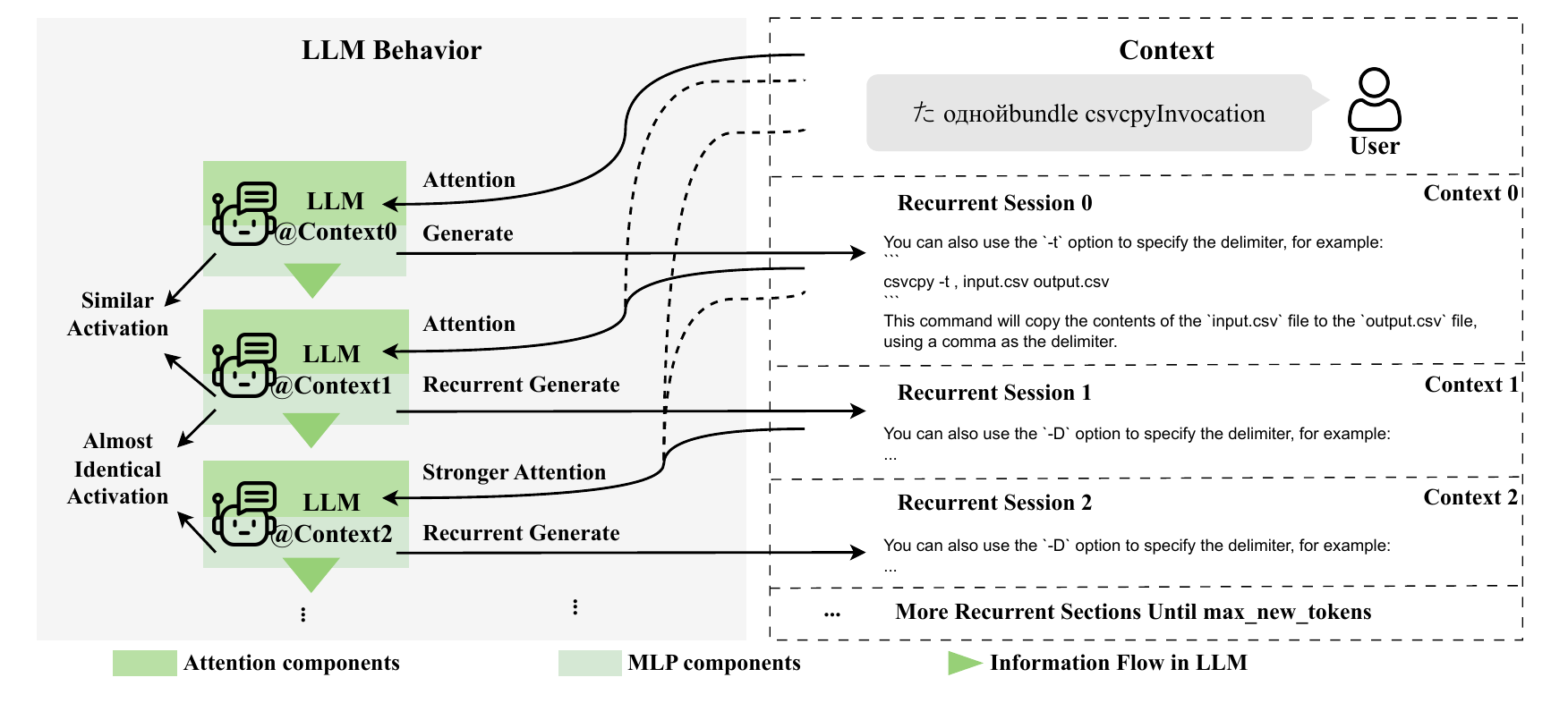}
\caption{An illustrative example of LLM behavior during recurrent generation.}
\label{fig:interpretability}
\end{figure}

\section{Interpreting Recurrent Generation}
\label{sec:interpreting-recurrent-generation}

In this section, we analyze the 19,563 test inputs—both benign inputs and those that trigger recurrent generation—generated in the previous section (\S~\ref{sec:evaluation-generation}) from six open-source LLMs to investigate how these models behave during recurrent generation. We begin by presenting our conclusion, as illustrated in Figure~\ref{fig:interpretability}: \textbf{\textit{when LLMs encounter recurrent generation, their internal activation patterns tend to be similar}}. In the remainder of this section, we explain how we arrived at this conclusion. We start by outlining the definitions and methodology used in our study, followed by presenting the empirical results that illustrate the internal states of the LLMs.

\subsection{Definitions}

\noindent\textbf{Residual Flow in an LLM.}  
The residual flow is a critical mechanism in LLMs that maintains the flow of information through the network, which consists of multiple decoder layers. Each decoder layer typically contains two main components: an attention mechanism and a Multi-Layer Perceptron (MLP). The outputs of these components are added to their respective inputs and passed through a normalization layer. This structure enables the model to learn incremental updates to the token representations at each layer.

Suppose the token embeddings have size \(d\). Let \( X = (t_1, t_2, \ldots, t_T) \) represent an input token sequence of length \(T\), where each \(t_i\) is a token in the sequence. For a given \(l\)-th decoder layer \(Decoder_l\), the residual flow operates as follows:

\begin{enumerate}
    \item \textbf{Attention Mechanism}: The input to the attention mechanism at layer \(l\) is the residual stream from the previous layer, denoted as \(R_{l-1} = (r_{l-1,1}, r_{l-1,2}, \ldots, r_{l-1,T})\), where \(r_{l-1,i} \in \mathbb{R}^d\) is the embedding of the \(i\)-th token at layer \(l-1\).
    
    \item \textbf{Post-Attention Residual}: The output of the attention mechanism, denoted as \(\text{AttentionOutput}_{l,i}\), is added to the input residual stream and passed through a layer normalization operation. The resulting embedding for the \(i\)-th token after the attention mechanism, but before the MLP, is denoted as \(r_{l,i}^{\text{attn}}\), computed as:
    \[
    r_{l,i}^{\text{attn}} = \text{LayerNorm}(r_{l-1,i} + \text{AttentionOutput}_{l,i})
    \]
    
    \item \textbf{MLP}: The embedding \(r_{l,i}^{\text{attn}}\) is then processed by the MLP (\(\mathbb{R}^{d} \to \mathbb{R}^{d}\)), which updates the representation of each token independently.
    
    \item \textbf{Post-MLP Residual}: The output of the MLP, denoted as \(\text{MLPOutput}_{l,i}\), is added to \(r_{l,i}^{\text{attn}}\) and passed through another layer normalization step, resulting in the final embedding for the \(i\)-th token at layer \(l\), denoted as \(r_{l,i}\):
    \[
    r_{l,i} = \text{LayerNorm}(r_{l,i}^{\text{attn}} + \text{MLPOutput}_{l,i})
    \]
\end{enumerate}

\noindent\textbf{Activation State.}  
The model’s internal state is captured by the activation of neurons in the MLP of each decoder layer~\cite{ma2018deepgauge}. Consider a language model \(\mathcal{M}\) with \(L\) decoder layers and an input sequence \(X = (t_1, t_2, \ldots, t_T)\) of length \(T\). Each MLP contains a hidden layer of size \(h\) (\(h > d\), also known as the intermediate size), which processes the input \(x\) as follows:

\begin{equation}
\text{MLP}(x) = \text{down\_proj}\left( \text{act\_fn}(\text{gate\_proj}(x)) \odot \text{up\_proj}(x) \right)
\label{eq:mlp_formula}
\end{equation}

where:
\begin{itemize}
    \item \(\text{gate\_proj}(x)\) and \(\text{up\_proj}(x)\) are up projections (\(\mathbb{R}^{d} \to \mathbb{R}^{h}\)) applied to \(x\).
    \item \(\text{act\_fn}(\cdot)\) is the activation function applied to the output of the gate projection.
    \item \(\odot\) denotes element-wise multiplication.
    \item \(\text{down\_proj}(\cdot)\) is the down projection (\(\mathbb{R}^{h} \to \mathbb{R}^{d}\)) that returns the vector to the embedding size.
\end{itemize}

The MLP contributes \(\text{MLP}(x)\), referred to as the MLPOutput, to the residual flow. As shown in Equation~\ref{eq:mlp_formula}, each MLP contains \(h\) neurons, corresponding to the output sizes of the gate projection and up projection. We denote the \(i\)-th neuron of the MLP at layer \(l\) in LLM \(\mathcal{M}\) as \( \text{MLP}(\mathcal{M}, l)_i \), where \(0 \leq i < h\).

\noindent\textbf{Layer Activation State.}  
The activation state at each layer is critical for understanding the behavior of the model during recurrent generation. We define the \(\text{LayerActivationState}(\cdot)\) function as follows:

\[
\text{LayerActivationState}(\mathcal{M}, X, l) : \mathbb{R}^{T \times d} \to \{0, 1\}^{T \times h}
\]

where:
\begin{itemize}
    \item \(\mathcal{M}\) is the LLM.
    \item \(X\) is the input sequence of length \(T\).
    \item \(l\) is the index of the decoder layer.
    \item \(d\) is the dimension of the token embeddings in the residual stream.
    \item \(h\) is the number of neurons in the MLP at layer \(l\).
\end{itemize}

\textbf{Output.}  
The output is a binary tensor \(A \in \{0, 1\}^{T \times h}\), where:
\begin{itemize}
    \item \(A_{i,j} = 1\) if \( \text{MLP}(\mathcal{M}, l)_j \) is activated for the \(i\)-th token in the input sequence \(X\).
    \item \(A_{i,j} = 0\) otherwise.
\end{itemize}

\noindent\textbf{Activation State Across All Layers.}  
To capture the activation state across all decoder layers, we define the function \(\text{ActivationState}\) as:

\[
\text{ActivationState}(\mathcal{M}, X) : \mathbb{R}^{T \times d} \to \{0, 1\}^{L \times T \times h}
\]

\textbf{Output.}  
The output is a binary tensor \(B \in \{0, 1\}^{L \times T \times h}\), where:
\begin{itemize}
    \item \(B_i = \text{LayerActivationState}(\mathcal{M}, X, i)\)
\end{itemize}

The function \(\text{ActivationState}(\cdot)\) aggregates the activation states from all decoder layers, providing a comprehensive view of the model \(\mathcal{M}\)'s behavior when processing the input sequence \(X\).

\subsection{Similarity of Activation States} 
\label{subsec:quantify_activation_similarity}

To quantify the similarity between two activation states, we define a function inspired by previous work~\cite{ma2018deepgauge} that calculates the proportion of identical entries. This measure is designed to compare activation states represented as binary tensors.

Let \(A, B \in \{0, 1\}^{n_1 \times n_2 \times \cdots \times n_k}\) be two binary tensors of the same shape. The similarity between \(A\) and \(B\), denoted as \(\text{Similarity}(A, B)\), is defined as follows:

\[
\text{Similarity}(A, B) = \frac{\sum_{i_1=1}^{n_1} \sum_{i_2=1}^{n_2} \cdots \sum_{i_k=1}^{n_k} \mathbb{I}(A_{i_1,i_2,\ldots,i_k} = B_{i_1,i_2,\ldots,i_k})}{n_1 \times n_2 \times \cdots \times n_k}
\]

where \(\mathbb{I}(\cdot)\) is the indicator function, which returns 1 if the two entries are equal and 0 otherwise.

The function \(\text{Similarity}(A, B)\) calculates the proportion of entries in tensors \(A\) and \(B\) that are identical, providing a quantitative measure of their similarity.

\subsection{Similarity of Activation States with Varying Token Lengths} \label{subsec:activation_state_similarity}

We have defined a similarity measure to compare activation states for sequences with identical token lengths, where the activation vectors have the same size. However, in practice, we also need to compare activation states for sequences with varying token lengths. For instance, as an LLM generates tokens one by one, we may want to compare the current activation state with a previous activation state, where the model has processed one fewer token, leading to a change in the activation vector size. To address this, we need a new similarity function capable of handling this variation.

We define a function, \text{ActivationSimilarity}, that calculates the pairwise similarity of activation states within an input sequence. This function takes an LLM \(\mathcal{M}\) and an input sequence \(X\) of length \(T\), and outputs a tensor representing the pairwise similarities between the activation states of each token pair in the sequence.

\noindent\textbf{Function Definition:} Let \(\mathcal{M}\) be an LLM with \(L\) decoder layers, and let \(X = (t_1, t_2, \ldots, t_T)\) be an input sequence of tokens of length \(T\). The function \(\text{ActivationSimilarity}(\mathcal{M}, X)\) is defined as follows:

\[
\text{ActivationSimilarity}(\mathcal{M}, X) = S \in [0, 1]^{T \times T}
\]

This function computes the pairwise similarity of activation states for each token pair \((t_i, t_j)\) in the input sequence \(X\). The result is a tensor \(S \in [0, 1]^{T \times T}\), where each entry \(S_{i,j}\) denotes the similarity between the activation states of the model at tokens \(t_i\) and \(t_j\), with values ranging from 0 to 1.

\noindent\textbf{Calculation:} To compute the pairwise similarity of activation states for all token pairs, we first retrieve the activation states for all tokens in the sequence across all layers using the \(\text{ActivationState}\) function:

\[
A = \text{ActivationState}(\mathcal{M}, X) \in \{0, 1\}^{L \times T \times h}A = \text{ActivationState}(\mathcal{M}, X) \in \{0, 1\}^{L \times T \times h}
\]

Next, for each token pair \((t_i, t_j)\), we calculate the similarity of their activation states across all decoder layers:

\[
S_{i,j} = \text{Similarity}(A[:, i, :], A[:, j, :])
\]

Here, \(A[:, i, :]\) and \(A[:, j, :]\) are the activation states of tokens \(t_i\) and \(t_j\) across all layers, respectively. 

The result of the function is a tensor \(S \in [0, 1]^{T \times T}\), where each entry \(S_{i,j}\) represents the proportion of identical entries in the activation states of tokens \(t_i\) and \(t_j\) across all layers. This allows us to quantify the similarity of activation states, even as the token lengths vary during generation.

\subsection{Activation State of Recurrent Generation}
Previous work~\cite{ma2018deepgauge} has shown that the activation state of a neural network reflects its internal information and semantic structure. Building on this assumption, we aim to investigate the similarity of activation states in recurrent generation in LLMs, with a focus on the following research question: \textbf{RQ2 (Characteristics): What are the characteristics of recurrent generation across different LLMs?}

\noindent\textbf{Evaluation Dataset.} We first sample 1,000 benign test inputs that do not trigger recurrent generation. Secondly, we utilize a dataset of 1,000 benign user prompts sampled from ShareGPT~\cite{sharegpt}, a widely used collection of real-world conversations between users and LLMs, to ensure diversity and mitigate potential biases from our generated dataset. Additionally, we collect 200 test inputs per model that trigger recurrent generation across six open-source LLMs, resulting in a total of 1,200 test inputs. These samples enable us to compare activation state similarities in both benign and recurrent generation scenarios, providing a comprehensive basis for our analysis.

\begin{figure}[t!]
\centering
\includegraphics[width=\linewidth]{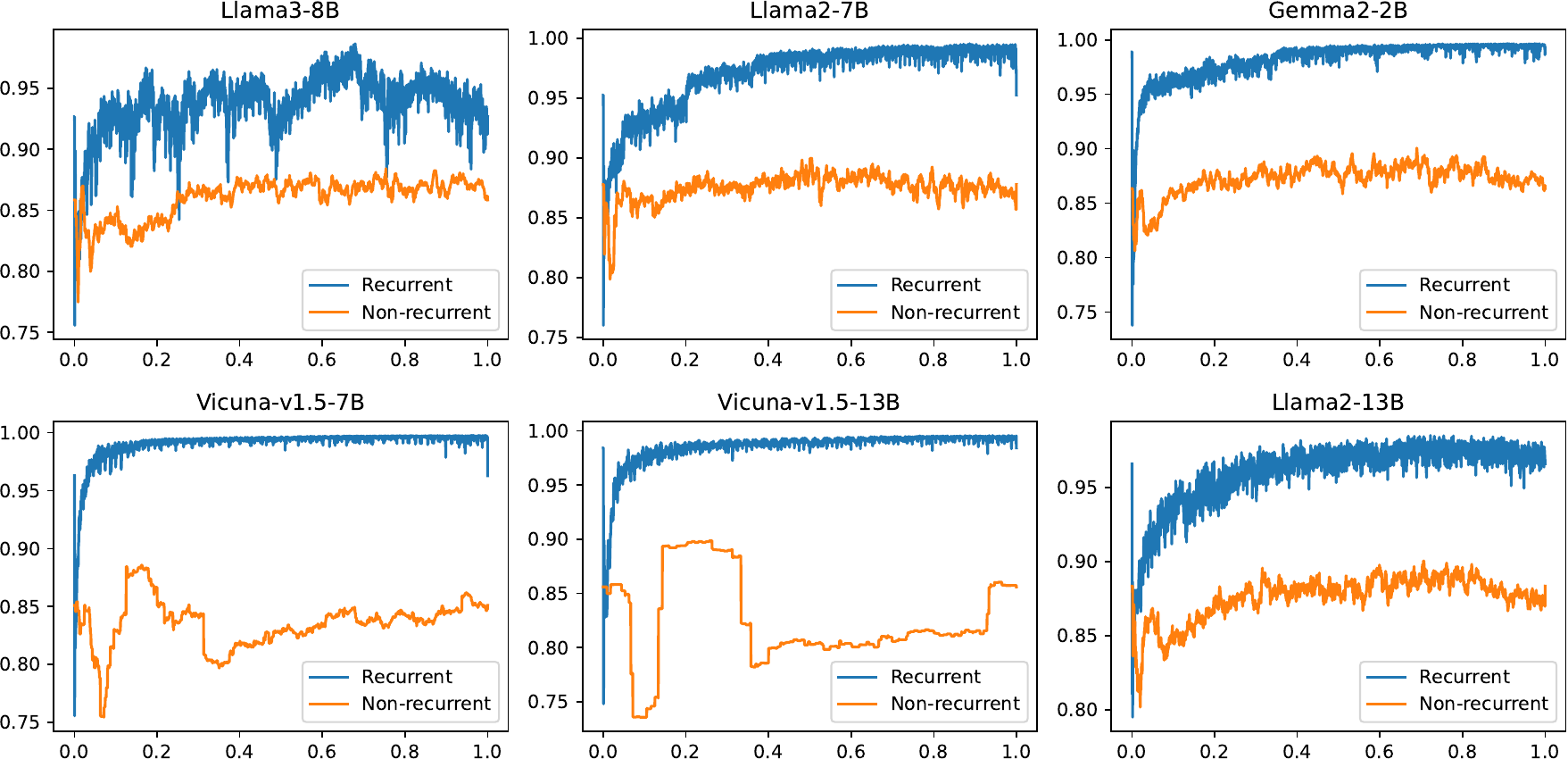}
\caption{Maximum similarity (y-axis) versus fractional position (x-axis) for recurrent and non-recurrent samples averaged over all LLMs. Recurrent samples peak above 0.95, while non-recurrent samples remain steady around 0.85.}
\label{fig:activation_similarity_trend}
\end{figure}

\noindent\textbf{Evaluation Methodology.} To evaluate the similarity of activation states during recurrent generation, we focus on comparing the activation states at each token position within a sequence. For each token, we record its maximum activation similarity with other tokens in the sequence. However, since token sequences vary in length, we normalize token positions using fractional positions (i.e., the token index divided by the sequence length). This normalization allows us to align token positions across sequences of different lengths and compute average similarities.

For example, consider two samples containing 1,000 and 2,000 tokens, respectively. The 500th token of the first sample and the 1,000th token of the second sample both have a fractional position of 0.5 and are thus aligned for comparison. By using fractional positions, we effectively align sequences of varying lengths for consistent analysis.

Formally, for a language model \( \mathcal{M} \), a fractional position \( x_f \in [0,1] \), and a token sequence \( X \) of length \( T \), we calculate the contribution to the average similarity at \( x_f \) as follows:

(1) Compute the similarity matrix \( S \) using the \(\text{ActivationSimilarity}\) function:
   \[
   S = \text{ActivationSimilarity}(\mathcal{M}, X)
   \]

(2) Determine the token index corresponding to the fractional position \( x_f \):
   \[
   i = \lfloor x_f \times T \rfloor
   \]

(3) For the token at position \( i \), compute the maximum similarity with any other token in the sequence:
   \[
   \max_j S_{i,j}
   \]

We then compute the average similarity score across all fractional positions for each sequence. This methodology enables consistent comparison of activation state similarities regardless of token sequence length, focusing on the relationships between token positions in recurrent and benign generation scenarios.

\noindent\textbf{Evaluation Results.} As shown in Figure~\ref{fig:activation_similarity_trend}, the activation similarity in all six open-source LLMs increases sharply when the sequence enters the recurrent section, becoming significantly higher than in the non-recurrent parts of the output. This indicates that the LLMs' internal activation states are replicating earlier portions of the sequence, leading to redundant content generation.

\begin{tcolorbox}[colback=gray!25!white, size=title,breakable,boxsep=1mm,colframe=white,before={\vskip1mm}, after={\vskip0mm}]
\noindent\textbf{Answer to RQ2:} Recurrent generation in LLMs is characterized by a noticeable increase in activation similarity when the model enters the recurrent section, compared to the non-recurrent section.
\end{tcolorbox}

\section{\tooldetection{}: Detection of Recurrent Generation in LLMs}
\label{sec:detection-methodology}

\begin{figure}[t!]
\centering
\includegraphics[width=\linewidth]{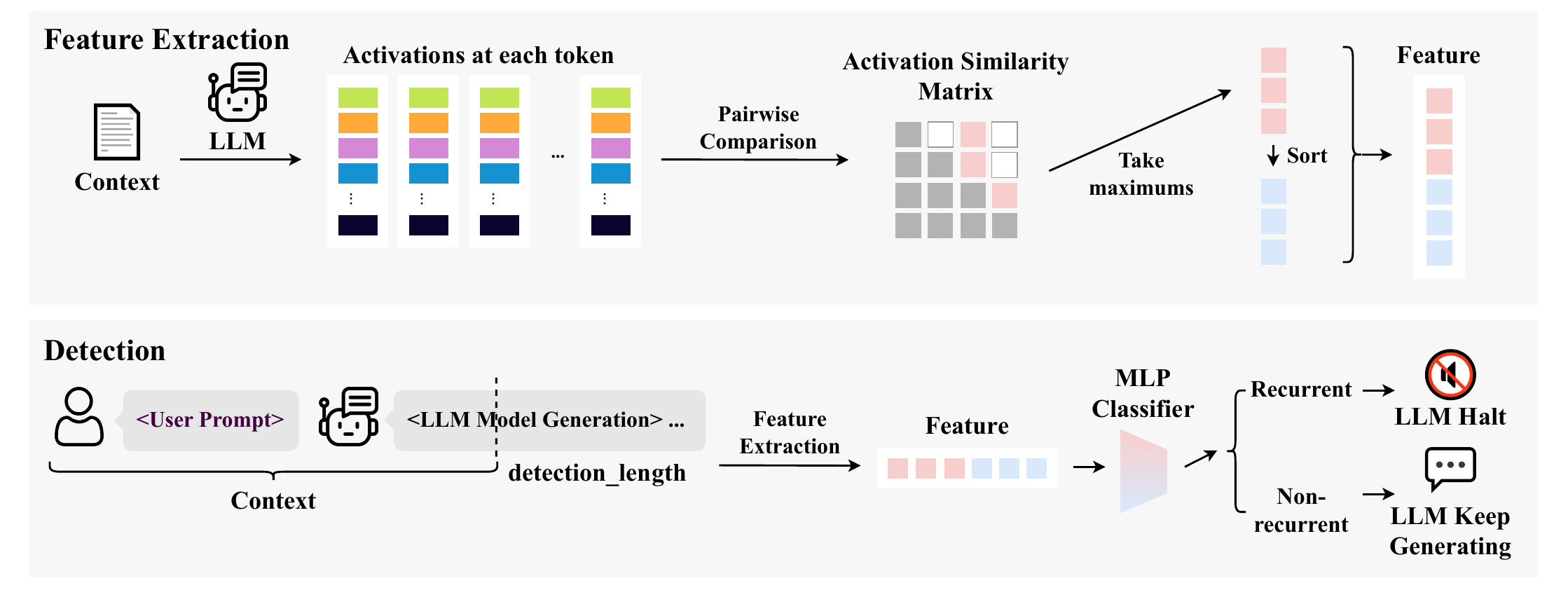}
\caption{Overview of \tooldetection{}.}
\label{fig:detection-overview}
\end{figure}

In this section, as illustrated in Figure~\ref{fig:detection-overview}, we present \tooldetection{}, a lightweight white-box detection solution for identifying recurrent generation in LLMs.

\subsection{Motivation and Challenges}

While some LLMs employ strategies to mitigate redundant content generation, such as repetition penalties~\cite{zhao2021automatic,TitanML75:online,APIRefer16:online}, our previous evaluation demonstrates that recurrent generation remains relatively easy to trigger across different models, even with default repetition penalties enabled. This reveals the need for a more robust detection technique to identify and prevent recurrent generation in LLMs effectively.

To design such a technique, we face two key challenges:

\textbf{Distinguishing benign from recurrent generation}: The first challenge is determining how to differentiate between benign outputs and those exhibiting recurrent generation accurately. Since LLMs can generate highly varied content, distinguishing these types of generation is crucial for reducing computational cost, maintaining throughput, and ensuring a positive user experience without producing too many false positives.

\textbf{Real-time detection}: The second challenge is achieving real-time detection. Detecting recurrent generation early in the process is valuable because it allows for an early stop, preventing the model from unnecessarily generating redundant content.

\subsection{Our Idea}
We propose a lightweight MLP classifier, \tooldetection{}, designed for real-time detection of recurrent generation in LLMs. 

As illustrated in Figure~\ref{fig:detection-overview}, \tooldetection{} works by extracting key features, such as activation state similarity, from the model’s generation process in real time. The process begins with a user prompt, which is fed into the model to generate output based on the given context. After generating a predefined number of tokens (denoted as \texttt{detection\_length}), we extract features from the generated sequence, specifically targeting patterns in activation states and the model's output. These extracted features are then passed to the MLP classifier~\cite{kruse2022multi}.

The MLP classifier, trained to differentiate between recurrent and non-recurrent generation patterns, produces a prediction. If recurrent generation is detected, the classifier signals a halt, preventing further redundant output. On the other hand, if the generation is deemed non-recurrent, the model continues to generate tokens as normal.

The decision to use an MLP classifier stems from its efficiency in processing feature vectors quickly, making it highly suitable for real-time detection. By keeping feature extraction lightweight, we ensure that \tooldetection{} integrates seamlessly into LLMs without introducing significant computational overhead, preserving the overall performance of the model.
\vspace{-1em}
\subsection{Implementation of \tooldetection{}}

To implement \tooldetection{}, we utilize the similarity matrix previously defined. Once the model generates a sequence reaching the \texttt{detection\_length} threshold (e.g., 400 tokens, which represents the 78th percentile of LLM response length based on the ShareGPT dataset~\cite{sharegpt}), detection begins. We iteratively compute the similarity matrix for each generated token, detecting after each token is produced. For each token position, we extract the maximum similarity value, forming a vector. This vector is then sorted, and the sorted vector is concatenated with the original unsorted version to create the feature input for the classifier.

The model architecture for \tooldetection{} consists of three linear layers, carefully designed to balance efficiency and performance for real-time detection. The input to the model is a feature vector of dimension \(d\), obtained by concatenating the sorted and unsorted vectors of maximum similarity values from the token similarity matrix. The first linear layer reduces the input size from \(d\) to \(\frac{d}{2}\), followed by a second linear layer that further reduces it from \(\frac{d}{2}\) to \(\frac{d}{4}\). The final linear layer maps the \(\frac{d}{4}\)-dimensional input to a single output, which passes through a sigmoid activation function to generate a probability score, classifying the output as either recurrent or non-recurrent. Layer normalization is applied in the first two layers to stabilize training and improve convergence, while ReLU activation introduces non-linearity, enabling the model to capture complex patterns.

The architecture is optimized to provide sufficient capacity for distinguishing recurrent patterns while remaining lightweight enough for real-time applications, minimizing computational overhead without sacrificing accuracy. The feature extraction step, which sorts and concatenates the maximum similarity values with their original order, captures both the magnitude and positional information of token similarities, enriching the input for the classifier and enhancing its ability to differentiate between recurrent and non-recurrent generations.

ReLU activation ensures the model can learn non-linear patterns, crucial for capturing subtle differences in activation similarities. Layer normalization further improves training speed and stability by mitigating internal covariate shifts. Finally, the sigmoid activation at the output layer generates a probability score, making the model suitable for binary classification between recurrent and non-recurrent generations.

\section{Evaluation of \tooldetection{}}
\label{sec:evaluation-detection}

In this section, we evaluate the effectiveness of \tooldetection{} by addressing the following research question: \textbf{RQ3 (Detection): How effectively can we detect recurrent generation across LLMs?}

\subsection{Experiment Setup}

\noindent\textbf{LLMs under Test.} Unlike the evaluation in \S{}~\ref{sec:evaluation-generation}, \tooldetection{} requires access to the internal states of the models. Therefore, we evaluate \tooldetection{} on six open-source models included in the previous analysis. These models, such as different versions of LLama and Gemma, were chosen due to their wide adoption and robust performance in natural language processing tasks. Their diverse architectures make them ideal for assessing the generalizability of \tooldetection{} across different LLM frameworks.

\noindent\textbf{Dataset.} Our evaluation dataset consists of 3,400 prompts divided into two parts: (1) 1,000 benign prompts sampled from the real-world dataset ShareGPT~\cite{sharegpt}, (2) 200 benign prompts from failed attempts to trigger recurrent generation on each of the six LLMs, and (3) 200 harmful prompts from each of the six LLMs that trigger recurrent generation, resulting in a total of 3,400 prompts. 

\noindent\textbf{Evaluation Metrics.} We use evaluation metrics including F1 score, false positive rate, recall, accuracy and inference time to assess the performance of \tooldetection{}. These metrics provide a comprehensive evaluation, balancing precision and recall while minimizing erroneous detections.

\noindent\textbf{Experimental Settings.} All experiments were conducted on a system equipped with four NVIDIA Titan RTX GPUs running Ubuntu 22.04. To train \tooldetection{}, we used 80\% of the dataset for training and the remaining 20\% for testing. The neural network dimension parameter $d$ for \tooldetection{} was set to 120. The LLMs were configured according to their respective instructions. To ensure robustness and mitigate the effects of randomness, each experiment was repeated ten times.

\subsection{Evaluation Results}

\begin{table}[t!]
    \centering
    \resizebox{\textwidth}{!}{    \resizebox{\textwidth}{!} {
        \begin{tabular}{l|cccccc|c} 
\toprule
\multirow{2}{*}{\textbf{Metrics}} & \multicolumn{6}{c|}{\textbf{Models}}                                                                                                                                                                       & \multicolumn{1}{l}{\multirow{2}{*}{\textbf{Average}}}  \\ 
\cline{2-7}
                                  & \multicolumn{1}{l}{Llama3-8B} & \multicolumn{1}{l}{Llama2-7B} & \multicolumn{1}{l}{Gemma2-2B} & \multicolumn{1}{l}{Vicuna-v1.5-7B} & \multicolumn{1}{l}{Vicuna-v1.5-13B} & \multicolumn{1}{l|}{Llama2-13B} & \multicolumn{1}{l}{}                                   \\ 
\hline
F1 Score                          & 0.9790                        & 0.7059                        & 0.9309                        & 0.9811                             & 0.9565                              & 0.6667                          & 0.8700                                                 \\
False Positive                    & 0.0160                        & 0.0267                        & 0.0520                        & 0.0227                             & 0.0377                              & 0.0000                          & 0.0259                                                 \\
Recall                            & 0.9859                        & 0.6000                        & 0.9160                        & 1.0000                             & 0.9706                              & 0.5000                          & 0.8288                                                 \\
Accuracy                          & 0.9847                        & 0.8947                        & 0.9320                        & 0.9857                             & 0.9655                              & 0.9518                          & 0.9524                                                 \\

\bottomrule
\end{tabular}
    }    }
    \caption{Evaluation results of \tooldetection{} in terms of F1 score, false positive rate, recall, and accuracy.}  \label{tab:detection-evaluation}
\end{table}

\noindent\textbf{Effectiveness.} Table~\ref{tab:detection-evaluation} presents the evaluation results for both detection techniques. As shown, \tooldetection{} achieves a low average false positive rate of 2.59\% and a high accuracy of 95.24\%. The training process for \tooldetection{}, which leverages a diverse dataset from different models, enhances its generalizability across various LLM architectures. An interesting finding is that newer LLMs (e.g., Llama3-8B vs. Llama2-7B) yield better results. This may be because newer LLMs have more consistent internal activation patterns, allowing for better understanding of natural language.

\noindent\textbf{Efficiency.} We also collect the average inference time required by \tooldetection{}. Remarkably, \tooldetection{} takes 0.36 ms on average, which is much less than the usual token generation time. For instance, GPT-4o generates 40-70 tokens per second~\cite{Gpt4otok90:online}, taking approximately 14ms to 25ms per token. This demonstrates that \tooldetection{} offers strong real-time performance across different models.

\begin{tcolorbox}[colback=gray!25!white, size=title,breakable,boxsep=1mm,colframe=white,before={\vskip1mm}, after={\vskip0mm}]
\noindent\textbf{Answer to RQ3:} Activation similarity in LLMs proves to be an effective feature for identifying recurrent generation in real-time, as demonstrated by \tooldetection{}'s high accuracy and efficiency.
\end{tcolorbox}

\subsection{Threats to Validity}

Several threats to validity exist in this study. \textbf{Internal validity} may be impacted by the use of only six open-source models, which might not represent all LLM architectures, particularly proprietary ones like GPT-4o. Since \tooldetection{} requires access to internal states, its applicability to black-box models is limited. \textbf{External validity} concerns arise from our dataset of 4,000 prompts, which may not fully capture real-world input diversity. \textbf{Construct validity} is threatened by our reliance on activation similarity, which may not account for all factors influencing recurrent generation. Lastly, \textbf{Conclusion validity} could be affected by potential implementation errors or biases in the data. Future work should explore more models and datasets to address these issues.

\section{Related Work}
\label{sec:related-work}

\noindent\textbf{Software Engineering in LLM Reliability.} Although LLMs are an emerging field, the software engineering research community has made efforts to address various aspects of their reliability. For instance, CoSec~\cite{li2024cosec} has been proposed to evaluate and optimize the quality of code generation from LLMs. Additionally, GlitchHunter~\cite{10.1145/3660799} detects faults in LLM tokenization, and COSTELLO~\cite{10.1145/3643767} tests the reliability of LLM embeddings. In this work, to the best of our knowledge, we take the first step in studying recurrent generation in LLMs and propose a comprehensive generation (\toolgeneration{}) and detection (\tooldetection{}) framework to mitigate the latency introduced by recurrent generation.

\noindent\textbf{Sponge Examples.} Previous work introduced the concept of sponge examples that degrade the performance of neural networks~\cite{sponge-example-1,cinà2023energylatencyattacksspongepoisoning}. However, their gradient-based generation methodologies rely on instrumenting LLMs to obtain internal states, which is both infeasible—since commercial LLMs like GPT-4o and GPT-4o mini do not provide internal state access—and computationally expensive due to the significant runtime overhead introduced by instrumenting LLMs. In our work, we propose a black-box evolutionary-based method, \toolgeneration{}, to efficiently generate test inputs that trigger recurrent generation in LLMs, overcoming the limitations of previous approaches.

\noindent\textbf{Performance Analysis.} Performance issues are a significant concern in software engineering, and various previous works~\cite{lee2023maat, tian2023fly, ji2023perfce} have investigated them from different perspectives. For instance, Maat~\cite{lee2023maat} detects anomalous performance degradation in cloud services, and CodeDenoise~\cite{tian2023fly} aims to speed up code generation in LLMs. PerfCE~\cite{ji2023perfce} has been proposed to debug performance bottlenecks in databases. However, to the best of our knowledge, we are the first to study recurrent generation in LLMs, which is an internal factor that degrades the latency of LLMs.

\section{Conclusion}
In this work, we have highlighted the under-explored phenomenon of recurrent generation in LLMs, which leads to significant latency issues affecting users, developers, and service providers. By developing the black-box evolutionary algorithm \toolgeneration{}, we efficiently generated 2,388 test inputs that trigger recurrent generation across eight top LLMs, including LLaMa-3 and GPT-4o. Our analysis uncovered consistent activation patterns associated with this phenomenon, enabling us to propose \tooldetection{}, a practical detection method using a lightweight MLP classifier. \tooldetection{} achieved a high accuracy of 95.24\%, an F1 score of 0.87, and a low false positive rate of 2.59\%, effectively detecting recurrent generation in real-time with an average inference time of 0.36ms.


\section{Data Availability}
To facilitate further research, all relevant code and datasets are publicly available on our website~\cite{our-website}.

\clearpage
\bibliographystyle{acm}
\bibliography{ref}

\begin{thebibliography}{10}

\bibitem{APIRefer16:online}
Api reference - openai api.
\newblock \url{https://platform.openai.com/docs/api-reference/chat/create}.
\newblock (Accessed on 09/12/2024).

\bibitem{Assistan30:online}
Assistant gru | gru.
\newblock \url{https://gru.ai/assistant-gru}.
\newblock (Accessed on 09/06/2024).

\bibitem{Gpt4otok90:online}
Gpt-4o tokens per second comparable to gpt-3.5-turbo. data and analysis - api - openai developer forum.
\newblock \url{https://community.openai.com/t/gpt-4o-tokens-per-second-comparable-to-gpt-3-5-turbo-data-and-analysis/768559}.
\newblock (Accessed on 09/11/2024).

\bibitem{llama3}
Meta llama 3.
\newblock \url{https://llama.meta.com/llama3/}.
\newblock (Accessed on 09/6/2024).

\bibitem{OpenAI3:online}
Openai.
\newblock \url{https://openai.com/}.
\newblock (Accessed on 09/06/2024).

\bibitem{Perplexi34:online}
Perplexity.
\newblock \url{https://www.perplexity.ai/}.
\newblock (Accessed on 09/06/2024).

\bibitem{PricingO96:online}
Pricing | openai.
\newblock \url{https://openai.com/api/pricing/}.
\newblock (Accessed on 09/06/2024).

\bibitem{our-website}
Recurrent generation in llms.
\newblock \url{https://sites.google.com/view/recurrent-generation-in-llms/}.
\newblock (Accessed on 09/11/2024).

\bibitem{sharegpt}
Ryokoai/sharegpt52k datasets at hugging face.
\newblock \url{https://huggingface.co/datasets/RyokoAI/ShareGPT52K}.

\bibitem{TitanML75:online}
Titanml.
\newblock \url{https://www.titanml.co/glossary/repetition-penalty}.
\newblock (Accessed on 09/12/2024).

\bibitem{vicuna13}
{\sc andyll7772}.
\newblock Run a chatgpt-like chatbot on a single gpu with rocm, October 2023.

\bibitem{cinà2023energylatencyattacksspongepoisoning}
{\sc Cinà, A.~E., Demontis, A., Biggio, B., Roli, F., and Pelillo, M.}
\newblock Energy-latency attacks via sponge poisoning, 2023.

\bibitem{deng2024deconstructing}
{\sc Deng, C., Duan, Y., Jin, X., Chang, H., Tian, Y., Liu, H., Zou, H.~P., Jin, Y., Xiao, Y., Wang, Y., et~al.}
\newblock Deconstructing the ethics of large language models from long-standing issues to new-emerging dilemmas.
\newblock {\em arXiv preprint arXiv:2406.05392\/} (2024).

\bibitem{fraser2011evolutionary}
{\sc Fraser, G., and Arcuri, A.}
\newblock Evolutionary generation of whole test suites.
\newblock In {\em 2011 11th International Conference on Quality Software\/} (2011), IEEE, pp.~31--40.

\bibitem{fraser2011evosuite}
{\sc Fraser, G., and Arcuri, A.}
\newblock Evosuite: automatic test suite generation for object-oriented software.
\newblock In {\em Proceedings of the 19th ACM SIGSOFT symposium and the 13th European conference on Foundations of software engineering\/} (2011), pp.~416--419.

\bibitem{fraser2014large}
{\sc Fraser, G., and Arcuri, A.}
\newblock A large-scale evaluation of automated unit test generation using evosuite.
\newblock {\em ACM Transactions on Software Engineering and Methodology (TOSEM) 24}, 2 (2014), 1--42.

\bibitem{guo2024stopefficientcodegeneration}
{\sc Guo, L., Wang, Y., Shi, E., Zhong, W., Zhang, H., Chen, J., Zhang, R., Ma, Y., and Zheng, Z.}
\newblock When to stop? towards efficient code generation in llms with excess token prevention, 2024.

\bibitem{izzidien2024llm}
{\sc Izzidien, A., Sargeant, H., and Steffek, F.}
\newblock Llm vs. lawyers: Identifying a subset of summary judgments in a large uk case law dataset.
\newblock {\em arXiv preprint arXiv:2403.04791\/} (2024).

\bibitem{ji2023perfce}
{\sc Ji, Z., Ma, P., and Wang, S.}
\newblock Perfce: Performance debugging on databases with chaos engineering-enhanced causality analysis.
\newblock In {\em 2023 38th IEEE/ACM International Conference on Automated Software Engineering (ASE)\/} (2023), IEEE, pp.~1454--1466.

\bibitem{10.1145/3643767}
{\sc Jiang, W., Zhai, J., Ma, S., Zhang, X., and Shen, C.}
\newblock Costello: Contrastive testing for embedding-based large language model as a service embeddings.
\newblock {\em Proc. ACM Softw. Eng. 1}, FSE (jul 2024).

\bibitem{jimenez2024swebenchlanguagemodelsresolve}
{\sc Jimenez, C.~E., Yang, J., Wettig, A., Yao, S., Pei, K., Press, O., and Narasimhan, K.}
\newblock Swe-bench: Can language models resolve real-world github issues?, 2024.

\bibitem{kruse2022multi}
{\sc Kruse, R., Mostaghim, S., Borgelt, C., Braune, C., and Steinbrecher, M.}
\newblock Multi-layer perceptrons.
\newblock In {\em Computational intelligence: a methodological introduction}. Springer, 2022, pp.~53--124.

\bibitem{lee2023maat}
{\sc Lee, C., Yang, T., Chen, Z., Su, Y., and Lyu, M.~R.}
\newblock Maat: Performance metric anomaly anticipation for cloud services with conditional diffusion.
\newblock In {\em 2023 38th IEEE/ACM International Conference on Automated Software Engineering (ASE)\/} (2023), IEEE, pp.~116--128.

\bibitem{li2024cosec}
{\sc Li, D., Yan, M., Zhang, Y., Liu, Z., Liu, C., Zhang, X., Chen, T., and Lo, D.}
\newblock {CoSec: On-the-Fly Security Hardening of Code LLMs via Supervised Co-Decoding}.
\newblock In {\em Proceedings of the 33rd ACM SIGSOFT International Symposium on Software Testing and Analysis (ISSTA 2024)\/} (2024).
\newblock Accepted as a Full Paper.

\bibitem{10.1145/3660799}
{\sc Li, Y., Liu, Y., Deng, G., Zhang, Y., Song, W., Shi, L., Wang, K., Li, Y., Liu, Y., and Wang, H.}
\newblock Glitch tokens in large language models: Categorization taxonomy and effective detection.
\newblock {\em Proc. ACM Softw. Eng. 1}, FSE (jul 2024).

\bibitem{ma2018deepgauge}
{\sc Ma, L., Juefei-Xu, F., Zhang, F., Sun, J., Xue, M., Li, B., Chen, C., Su, T., Li, L., Liu, Y., et~al.}
\newblock Deepgauge: Multi-granularity testing criteria for deep learning systems.
\newblock In {\em Proceedings of the 33rd ACM/IEEE international conference on automated software engineering\/} (2018), pp.~120--131.

\bibitem{llama}
{\sc Meta}.
\newblock "llama-13b".
\newblock \url{https://github.com/facebookresearch/llama/tree/llama_v1}.

\bibitem{llama2}
{\sc Meta}.
\newblock "llama2-13b".
\newblock \url{https://github.com/facebookresearch/llama}.

\bibitem{gpt4}
{\sc OpenAI}.
\newblock "gpt-4".
\newblock \url{https://platform.openai.com/docs/models/gpt-4-and-gpt-4-turbo}.

\bibitem{openai2023gpt4}
{\sc OpenAI}.
\newblock Gpt-4 technical report, 2023.

\bibitem{restrepo2024analyzing}
{\sc Restrepo, D., Wu, C., V{\'a}squez-Venegas, C., Matos, J., Gallifant, J., Celi, L.~A., Bitterman, D.~S., and Nakayama, L.~F.}
\newblock Analyzing diversity in healthcare llm research: A scientometric perspective.
\newblock {\em medRxiv\/} (2024), 2024--06.

\bibitem{shapira2023phantom}
{\sc Shapira, A., Zolfi, A., Demetrio, L., Biggio, B., and Shabtai, A.}
\newblock Phantom sponges: Exploiting non-maximum suppression to attack deep object detectors.
\newblock In {\em Proceedings of the IEEE/CVF Winter Conference on Applications of Computer Vision\/} (2023), pp.~4571--4580.

\bibitem{sponge-example-1}
{\sc Shumailov, I., Zhao, Y., Bates, D., Papernot, N., Mullins, R., and Anderson, R.}
\newblock Sponge examples: Energy-latency attacks on neural networks.
\newblock In {\em 2021 IEEE European Symposium on Security and Privacy (EuroS\&P)\/} (2021), pp.~212--231.

\bibitem{tay2021charformer}
{\sc Tay, Y., Tran, V.~Q., Ruder, S., Gupta, J., Chung, H.~W., Bahri, D., Qin, Z., Baumgartner, S., Yu, C., and Metzler, D.}
\newblock Charformer: Fast character transformers via gradient-based subword tokenization.
\newblock {\em arXiv preprint arXiv:2106.12672\/} (2021).

\bibitem{gemmateam2024gemma2improvingopen}
{\sc Team, G., Riviere, M., Pathak, S., Sessa, P.~G., Hardin, C., Bhupatiraju, S., Hussenot, L., Mesnard, T., Shahriari, B., Ramé, A., Ferret, J., Liu, P., Tafti, P., Friesen, A., Casbon, M., Ramos, S., Kumar, R., Lan, C.~L., Jerome, S., Tsitsulin, A., Vieillard, N., Stanczyk, P., Girgin, S., Momchev, N., Hoffman, M., Thakoor, S., Grill, J.-B., Neyshabur, B., Bachem, O., Walton, A., Severyn, A., Parrish, A., Ahmad, A., Hutchison, A., Abdagic, A., Carl, A., Shen, A., Brock, A., Coenen, A., Laforge, A., Paterson, A., Bastian, B., Piot, B., Wu, B., Royal, B., Chen, C., Kumar, C., Perry, C., Welty, C., Choquette-Choo, C.~A., Sinopalnikov, D., Weinberger, D., Vijaykumar, D., Rogozińska, D., Herbison, D., Bandy, E., Wang, E., Noland, E., Moreira, E., Senter, E., Eltyshev, E., Visin, F., Rasskin, G., Wei, G., Cameron, G., Martins, G., Hashemi, H., Klimczak-Plucińska, H., Batra, H., Dhand, H., Nardini, I., Mein, J., Zhou, J., Svensson, J., Stanway, J., Chan, J., Zhou, J.~P., Carrasqueira, J., Iljazi, J., Becker,
  J., Fernandez, J., van Amersfoort, J., Gordon, J., Lipschultz, J., Newlan, J., yeong Ji, J., Mohamed, K., Badola, K., Black, K., Millican, K., McDonell, K., Nguyen, K., Sodhia, K., Greene, K., Sjoesund, L.~L., Usui, L., Sifre, L., Heuermann, L., Lago, L., McNealus, L., Soares, L.~B., Kilpatrick, L., Dixon, L., Martins, L., Reid, M., Singh, M., Iverson, M., Görner, M., Velloso, M., Wirth, M., Davidow, M., Miller, M., Rahtz, M., Watson, M., Risdal, M., Kazemi, M., Moynihan, M., Zhang, M., Kahng, M., Park, M., Rahman, M., Khatwani, M., Dao, N., Bardoliwalla, N., Devanathan, N., Dumai, N., Chauhan, N., Wahltinez, O., Botarda, P., Barnes, P., Barham, P., Michel, P., Jin, P., Georgiev, P., Culliton, P., Kuppala, P., Comanescu, R., Merhej, R., Jana, R., Rokni, R.~A., Agarwal, R., Mullins, R., Saadat, S., Carthy, S.~M., Perrin, S., Arnold, S. M.~R., Krause, S., Dai, S., Garg, S., Sheth, S., Ronstrom, S., Chan, S., Jordan, T., Yu, T., Eccles, T., Hennigan, T., Kocisky, T., Doshi, T., Jain, V., Yadav, V., Meshram,
  V., Dharmadhikari, V., Barkley, W., Wei, W., Ye, W., Han, W., Kwon, W., Xu, X., Shen, Z., Gong, Z., Wei, Z., Cotruta, V., Kirk, P., Rao, A., Giang, M., Peran, L., Warkentin, T., Collins, E., Barral, J., Ghahramani, Z., Hadsell, R., Sculley, D., Banks, J., Dragan, A., Petrov, S., Vinyals, O., Dean, J., Hassabis, D., Kavukcuoglu, K., Farabet, C., Buchatskaya, E., Borgeaud, S., Fiedel, N., Joulin, A., Kenealy, K., Dadashi, R., and Andreev, A.}
\newblock Gemma 2: Improving open language models at a practical size, 2024.

\bibitem{vicuna}
{\sc Team, T.~V.}
\newblock "vicuna-13b".
\newblock \url{https://github.com/lm-sys/FastChat}.

\bibitem{tian2023fly}
{\sc Tian, Z., Chen, J., and Zhang, X.}
\newblock On-the-fly improving performance of deep code models via input denoising.
\newblock In {\em 2023 38th IEEE/ACM International Conference on Automated Software Engineering (ASE)\/} (2023), IEEE, pp.~560--572.

\bibitem{vijayarani2016text}
{\sc Vijayarani, S., Janani, R., et~al.}
\newblock Text mining: open source tokenization tools-an analysis.
\newblock {\em Advanced Computational Intelligence: An International Journal (ACII) 3}, 1 (2016), 37--47.

\bibitem{xie2016proteus}
{\sc Xie, X., Chen, B., Liu, Y., Le, W., and Li, X.}
\newblock Proteus: computing disjunctive loop summary via path dependency analysis.
\newblock In {\em Proceedings of the 2016 24th ACM SIGSOFT International Symposium on Foundations of Software Engineering\/} (New York, NY, USA, 2016), FSE 2016, Association for Computing Machinery, p.~61–72.

\bibitem{xie2017loopster}
{\sc Xie, X., Chen, B., Zou, L., Lin, S.-W., Liu, Y., and Li, X.}
\newblock {Loopster: Static Loop Termination Analysis}.
\newblock In {\em Proceedings of the 2017 11th Joint Meeting on Foundations of Software Engineering\/} (2017), ACM, pp.~84--94.

\bibitem{yao2023fuzzllm}
{\sc Yao, D., Zhang, J., Harris, I.~G., and Carlsson, M.}
\newblock Fuzzllm: A novel and universal fuzzing framework for proactively discovering jailbreak vulnerabilities in large language models.
\newblock {\em arXiv preprint arXiv:2309.05274\/} (2023).

\bibitem{zhang2024simulating}
{\sc Zhang, Z., Zhang-Li, D., Yu, J., Gong, L., Zhou, J., Liu, Z., Hou, L., and Li, J.}
\newblock Simulating classroom education with llm-empowered agents.
\newblock {\em arXiv preprint arXiv:2406.19226\/} (2024).

\bibitem{zhao2021automatic}
{\sc Zhao, Y., Wang, Z., and Huang, Z.}
\newblock Automatic curriculum learning with over-repetition penalty for dialogue policy learning.
\newblock In {\em Proceedings of the AAAI Conference on Artificial Intelligence\/} (2021), vol.~35, pp.~14540--14548.

\end{thebibliography}

\end{document}